%Paper: hep-ph/9206254
%From: "Stephen Naculich" <naculich@casa.pha.jhu.edu>
%Date: 26 Jun 92 12:29:00 EST

\input phyzzx
\overfullrule=0pt
%========================DEFINITIONS===================================
\def\sss{\scriptscriptstyle }
\def\ra{\rightarrow}
\def\d{{\rm d}}
\def\M {{\cal M}}
\def\ON{ O(N) }
\def\cutoff{ \Lambda_c }
\def\hats{{\tilde s}}
\def\hatcutoff{{\tilde\cutoff}}
\def\htau{{\tilde \tau}}
\def\LWW{ L_{\sss WW} }
\def\func{ F }

\nopubblock
\line{\hfil JHU-TIPAC-920017}
\line{\hfil June, 1992}
\titlepage
\title{ Can the Electroweak Symmetry-breaking Sector Be Hidden? }
\author{ S.~G.~Naculich \ \ and\ \ C.--P. Yuan}
\address{
Department of Physics and Astronomy \break
The Johns Hopkins University  \break
Baltimore, MD  21218 }

\abstract{
In a recent paper, Chivukula and Golden
claimed that the electroweak symmetry--breaking
sector could be hidden if there were
many inelastic channels in the
longitudinal $WW$ scattering process.
They presented a model in which the $W$'s
couple to pseudo--Goldstone bosons,
which may be difficult to detect experimentally.
Because of these inelastic channels,
the $WW$ interactions
do not become strong in the TeV region.
We demonstrate that,
despite the reduced $WW$ elastic amplitudes
in this model,
the total event rate
($\sim 5000$ extra longitudinal
$W^+W^-$ pairs produced in one standard SSC year)
does not decrease
with an increasing number of inelastic channels,
and is roughly the same as in a model
with a broad high--energy resonance
and no inelastic channels.
}

\endpage
%=========================Body of paper=======================

\REF\nolose{
M. S. Chanowitz and M. K. Gaillard, Nucl. Phys. {\bf B261}, 379 (1985);
\nextline
M. S. Chanowitz, Ann. Rev. Nucl. Part. Sci. {\bf 38}, 323 (1988).}
The ``no--lose theorem'' states that
if light Higgs bosons do not exist,
elastic longitudinal $W$ scattering
becomes strong at or above 1 TeV,
and that the new strong interactions
can be detected by observing
$WW$ scattering via leptonic decays of $W$'s.\refmark{\nolose}\
(We use $W$ to denote either the $W^\pm$ or $Z^0$ boson.)
In a recent paper,\Ref\CG{
R. S. Chivukula and M. Golden, Phys. Lett. {\bf B267}, 233 (1991). }
Chivukula and Golden have argued that
the ``no--lose theorem'' breaks down
if there are many inelastic channels
into which the $W$'s can scatter.
They presented a toy model in which the $W$'s
couple to a large number of pseudo--Goldstone bosons,
which may be difficult to detect experimentally.
Because of the large number of inelastic channels,
there are light resonances in
the elastic $WW$ scattering amplitudes,
which are too broad to be discernible as peaks.
Moreover,
the growth of the elastic scattering amplitudes
is cut off at the scale of the light resonances,
so the $WW$ interactions
do not become strong in the TeV region.
They conclude that,
unless the pseudo--Goldstone bosons themselves can be observed,
the electroweak--symmetry breaking sector will remain hidden.

\REF\ONmodel{
S.~Coleman, R.~Jackiw, and H.~Politzer, Phys. Rev. {\bf D10}, 2491  (1974);
\nextline
L.~Abbott, J.~Kang, and H.~Schnitzer, Phys. Rev. {\bf D13}, 2212 (1976);
\nextline
W.~Bardeen and M.~Moshe, Phys. Rev. {\bf D28}, 1372 (1983).}
\REF\EW{
M. B. Einhorn, Nucl. Phys. {\bf B246}, 75 (1984);
\nextline
M. B. Einhorn and D. N. Williams, Phys. Lett. {\bf 211B}, 457 (1988).}
In this note,
we point out that
the total event rate for elastic $WW$ scattering
in the model of ref.~\CG~does not decrease as the number
of inelastic channels increases.
The elastic amplitude is smaller,
but the resonance is at a lower energy,
where the parton luminosity is greater.
We give a simple scaling argument to show that
the total rate
($\sim 5000$ $W^+W^-$ pairs and
$\sim 2500$ $Z^0Z^0$ pairs
produced in one standard SSC year)
is roughly independent of the
number of inelastic channels,
and is about the same as that
in the standard $O(4)$ model\refmark{\ONmodel,\EW}
with no inelastic channels
and a broad TeV scale resonance.
We also briefly comment on possible methods
to detect the signal.

The model presented in ref.~\CG~to
demonstrate the possible effects
of inelastic channels in the electroweak sector
contains both exact Goldstone bosons
and a large number of pseudo--Goldstone bosons.
This model possesses an approximate $O(j+n)$ symmetry
which is explicitly broken to $O(j) \times O(n)$.
The exact $O(j)$ symmetry is spontaneously broken to $O(j-1)$,
yielding $j-1$ massless Goldstone bosons, $\phi$,
and one massive scalar boson.
The $O(n)$ symmetry remains unbroken,
and there are $n$ degenerate pseudo--Goldstone bosons,
$\psi$, with mass $m_\psi$.

To use this model to describe the scattering of longitudinal $W$'s,
one applies the equivalence theorem,
replacing the longitudinal $W$ with its corresponding Goldstone boson $\phi$
in the $S$--matrix.
This equivalence holds only when $ E_W \gg M_W $,
where $E_W$ is the energy of the $W$ boson
in the $WW$ center--of--mass frame.
Therefore, strictly speaking, one should not use this model
to describe $WW$ scattering when the invariant mass $M_{WW}$
is less than a couple of times the mass threshold $2 M_W$.
The amplitudes for the scattering of
longitudinal $W$'s are given by\Ref\kane{
G.~L.~Kane and C.--P.~Yuan, Phys.~Rev. {\bf D40}, 2231 (1989).
In this reference, eq.~(2.16) should be corrected in accord with eq.~(1)
of this paper. The effect of this correction is that
the $W^-W^+$ event rate discussed in this reference
for the $O(4)$ model should be multiplied by $\sim 2$ to 2.5. }
$$\eqalign{
\M (Z^0Z^0 \to W^-W^+) & = A(s,t,u), \cr
\M (W^-W^+ \to Z^0Z^0) & = A(s,t,u), \cr
\M (W^-W^+ \to W^-W^+) & = A(s,t,u)+A(t,s,u), \cr
\M (Z^0Z^0 \to Z^0Z^0) & = A(s,t,u)+A(t,s,u)+A(u,t,s), \cr
\M (W^\pm Z^0 \to W^\pm Z^0) & = A(t,s,u), \cr
\M (W^\pm W^\pm \to W^\pm W^\pm) & =A(t,s,u)+A(u,t,s). \cr}
\eqn\scattamp
$$
In the models we consider in this note,
$ A(s,t,u) = A(s)$ depends only on $s$.

Before turning to the model of ref.~\CG,
we recall some relevant features of
the $\ON$ model.\refmark{\ONmodel,\EW}
To leading order in $1/N$,
with the parameters $v$ and $\cutoff$ held fixed as $N \to \infty$,
the amplitude $A(s)$ in the $\ON$ model is given by
$$
A(s) = {s\over N}
\left\{
v^2 - { s \over  32 \pi^2 }
\left[ \ln \left({e\cutoff^2 \over  |s|} \right) + i \pi \Theta(s) \right]
\right\}^{-1}.
\eqn\ONamp
$$
Here $\Theta(s>0)=1$ and $\Theta(s<0)=0$,
and $\cutoff$ is the cut--off
scale of the theory,\Ref\cut{
Our $\cutoff$ is $\sqrt{e}$ times the one defined in ref.~\EW.}
related to the tachyon mass $\mu_t$ through
$$
\cutoff
={\mu_t \over \sqrt{e}} \exp\left( {-16 \pi^2 v^2 \over \mu_t^2}
\right).
\eqn\ONcutoff
$$
(For $\mu_t^2 \gg v^2$, the tachyon and cut--off scales
are roughly the same, $\cutoff \simeq \mu_t / \sqrt{e}$.)
Because of the presence of the tachyon,
the $\ON$ model must be regarded as an effective theory,
valid only at energy scales well below $\mu_t$.
With $v$, $\cutoff$, and $s$ held fixed,
the amplitude \ONamp~evidently scales as $1/N$.

To extract physical predictions from the $\ON$ model,
one must set $N$ equal to some finite value;
$ N=4 $ corresponds to the electroweak sector with
its three Goldstone bosons and one massive Higgs boson.
To ensure that low--energy theorems for the scattering
amplitudes are satisfied,
one must then set $v = f/\sqrt{N} $,
where $ f = 250$ GeV characterizes the symmetry--breaking scale.
The amplitude is then given by
$$
A(s) = s
 \left\{
f^2 - { sN \over  32 \pi^2 }
\left[ \ln \left({e\cutoff^2 \over  |s|} \right) + i \pi \Theta(s) \right]
\right\}^{-1}.
\eqn\ONamptwo
$$
With $f$ held fixed,
the scaling property of the amplitude differs slightly from
that described above;
$A(s)$ scales as $1/N$,
but only if we simultaneously scale $s$ with $1/N$
and $\cutoff$ with $1/\sqrt{N}$.
In other words,
$$
A(s) = {1\over N} \func(\hats,\hatcutoff),
{}~~~\quad \hats= Ns,
{}~~~\quad \hatcutoff=\sqrt{N}\cutoff,
\eqn\ONscaling
$$
where $\func(\hats,\hatcutoff)$ only depends on $N$
through $\hats$ and $\hatcutoff$.

To locate resonances in the
scattering amplitudes \scattamp,
we look for the position of the (complex) pole of $ A(s) $
as a function of the parameters of the theory.
The position of the pole $s$ can be parametrized by its
``mass'' $m$ and ``width'' $\Gamma$
through the relation $ s = (m -  {i\over 2} \Gamma)^2 $,
though we should not take these terms literally when $\Gamma$ is
comparable to $m$.
The pole traces out a curve in the $s$--plane
as $\mu_t$ is varied.
When $\mu_t$ is very large,
the real and imaginary parts of the pole are both small,
corresponding to a light, narrow resonance.
As $\mu_t$ decreases,
Re$(s)$ increases,
reaches a maximum
and then begins to decrease,
while Im$(s)$ continues to increase.
We refer to the pole position with maximum Re$(s)$
as the ``heaviest'' resonance.
This resonance is very broad,
with $\Gamma$ roughly equal to $m$.
In the $O(4)$ model,
the ``heaviest'' resonance is found\refmark\EW
to have ``mass'' $m= 845$ GeV and ``width'' $\Gamma =  640 $ GeV,
and corresponds to a cut--off $\cutoff = 4.9 $ TeV
and tachyon mass $\mu_t = 8.4 $ TeV.
{}From the scaling property of eq.~\ONamptwo,
the values of $m$ and $\Gamma$
corresponding to the heaviest resonance
for the $\ON$ model
are $\sqrt{4\over N}$
times those for the $O(4)$ model;
the cut--off and tachyon mass scale in the same way.

We now turn to the $O(j)\times O(n)$ model of ref.~\CG.
The amplitudes are calculated
in the limit $j$, $n \to  \infty$ with the ratio
$j/n$ fixed;
only diagrams which contribute to leading order in $1/(j+n)$
are included.
The amplitude $A(s)$ is given by
$$
 A(s) = s \left\{
f^2
 - {sj \over 32 \pi^2}
\left[ \ln\left(  e \cutoff^2 \over |s| \right) + i \pi\Theta(s) \right]
- { sn \over 32 \pi^2}
\left[ \ln\left( \cutoff^2 \over e m_\psi^2 \right) -F_2(s,m_\psi) \right]
\right\}^{-1},
\eqn\OJONamp
$$
where
$$
\eqalign{
F_2(s,m)&
=  -~2+~\sqrt{ 1 - {4m^2 \over s} }
{}~\ln\left(
{ \sqrt{4m^2-s} + \sqrt{-s}  \over
  \sqrt{4m^2-s} - \sqrt{-s}  } \right)
\quad \quad \quad {\rm for} \quad s<0,
\cr
F_2(s,m)& =-~2+2~\sqrt{ -1 + {4m^2 \over s}  } ~{\rm arctan}
\sqrt{ {s \over 4m^2-s} }
\quad \quad \quad \quad \quad {\rm for} \quad 0<s<4m^2, \cr
F_2(s,m)& =-~2+\sqrt{ 1 - {4m^2 \over s}}
\left[ \ln\left(
 { \sqrt{s} + \sqrt{s-4m^2} \over   \sqrt{s} -  \sqrt{s-4m^2} }
\right) - i  \pi\right]  \quad {\rm for} \quad s>4m^2, \cr}
\eqn\Ftwo
$$
and the cut--off $\cutoff$ (equal to $M$ of ref.~\CG)
is related to the tachyon scale by
$$
\cutoff
={\mu_t \over \sqrt{e}}
\exp\left\{ {-16 \pi^2 f^2 \over (j+n) \mu_t^2}
+{n\over 2(j+n)} \left[
  \ln \left( m_\psi^2 \over \mu_t^2 \right)
- \sqrt{1 + {4m_\psi^2 \over \mu_t^2}}
\ln \left(    \sqrt{ \mu_t^2+4m_\psi^2 } - \mu_t
         \over\sqrt{ \mu_t^2+4m_\psi^2 } +  \mu_t\right)
        \right] \right\}.
\eqn\OJONcutoff
$$
Qualitatively speaking,
for center--of--mass energies
well below the $\psi$ mass threshold,
$s \ll 4m_\psi^2$,
the $O(j)\times O(n)$ model
behaves like the $O(j)$ model;
the pseudo--Goldstone bosons play little role.
On the other hand,
well above the threshold,
$s \gg 4m_\psi^2$,
the $O(j)\times O(n)$ model
behaves like the $O(j+n)$ model.

Having obtained the amplitude \OJONamp,
one sets
$f=250$ GeV and $j=4$;
the exact Goldstone bosons in this model
correspond to the longitudinal $W$'s.
Three independent parameters now specify the model:
the number of pseudo--Goldstone bosons $n$,
their mass $m_\psi$,
and the tachyon mass $\mu_t$.
(Again, the model is only valid
at energy scales well below the tachyon mass.)

We now compare the total event rates in the
$O(4)\times O(n)$ model for different values of $n$.
As in the $\ON$ model,
the amplitude \OJONamp~has a complex pole,
whose real part increases and then decreases
as $\mu_t$ varies.
We choose the parameter $m_\psi$
so that the resonance is well above the
the pseudo--Goldstone mass threshold,
where the model essentially behaves
like the $O(4+n)$ model.
Thus, using the scaling behavior described earlier,
the ``heaviest'' resonance
of the $O(4)\times O(n)$ model
has $m \simeq \sqrt{4\over 4+ n}  \times 845  $ GeV
and $\Gamma\simeq \sqrt{4\over 4+n} \times 640  $  GeV;
the corresponding tachyon mass and cut--off
also scale as $ \sqrt{4\over 4+n}  $
relative to the $O(4)$ model.
We choose the tachyon mass $\mu_t$ for each value of $n$
to correspond to the ``heaviest'' resonance
of that model.
Note that as $n$ increases,
the resonance moves
to smaller mass $m$;
the width to mass ratio of the resonance
is of course independent of $n$.

Above the pseudo--Goldstone mass threshold,
where the model behaves like the $O(4+n)$ model,
the amplitude \OJONamp~has the scaling property
$$
A (s) = {1 \over 4+n} \func(\hats), ~~~\quad \hats= (4+n)s,
\eqn\OJONscaling
$$
This follows from eq.~\ONscaling~because
the cut--off $\cutoff$ for the heaviest resonance
scales as $1/\sqrt{4+n}$,
and so $\hatcutoff$ is independent of $n$.
Since the amplitude \OJONscaling~scales as $1/(4+n)$,
it would seem that the scattering rate
becomes smaller as $n$ increases.
On the other hand, for larger $n$,
the resonance occurs at lower invariant mass,
where the $WW$ parton luminosity $\LWW$ is higher.
We now show that the two effects cancel each other.

\REF\effwa{
M.~Chanowitz and M.~K.~Gaillard, Phys.~Lett.~{\bf B142}, 85 (1984);
\nextline
G.~Kane, W.~Repko, and W.~Rolnick, Phys.~Lett.~{\bf B148}, 367 (1984).}
\REF\effwb{S.~Dawson, Nucl.~Phys.~{\bf B249}, 42 (1985).}
In the effective--$W$ approximation\refmark{\effwa,\effwb}
we have
$$
\eqalign{
\sigma_{\sss pp \ra WW \ra WW} (S)
&
= \int_{\tau_{\rm min}}^1 \d \tau {\d \LWW \over \d \tau}
\sigma_{\sss WW \ra WW} (\tau S),
\cr
\sigma_{\sss WW \ra WW} (\tau S)
&
= \int \d \Omega {1\over 2 \tau S} \left| \M(\tau S) \right|^2 ,
\cr}
\eqn\eventrate
$$
where $\sqrt{S}$ is the center--of--mass energy of the $pp$ collider,
$\sqrt{s} = \sqrt{\tau S}$ is the invariant mass of the $WW$ pair,
$\tau_{\rm min} = 4M_W^2/S$,
and $\d \Omega$ integrates
over the direction of
the out--going $W$ in the $WW$ center--of--mass frame.
The parton luminosity
$ \left( \d \LWW / \d \tau \right) $
scales\Ref\luminosity{Based on fig. 5 of ref.~\effwb.}
approximately as $1/\tau^2$
for $M_{WW} \lsim 1$ TeV at the SSC.
By rewriting eq.~\eventrate~ in
terms of $\htau = (4+n)\tau$,
using eq.~\OJONscaling,
we find that $\sigma_{\sss pp \ra WW \ra WW} (S)$
is actually independent of $n$.
Thus we conclude that,
although the amplitude decreases
as $n$ increases,
the total elastic event rate stays the same.

To see how large the event rate actually is,
we choose $n=8$ and $m_\psi=125$ GeV.
We expect from our scaling arguments
that the heaviest resonance for $n=8$ will have
$m \simeq \sqrt{4\over 12}  \times 845  = 490 $ GeV
and $\Gamma\simeq \sqrt{4\over 12}  \times 640  = 370 $  GeV.
Indeed, using eq.~\OJONamp~explicitly,
we find that the heaviest resonance
occurs for $m = 485 $ GeV and $\Gamma = 350 $ GeV,
corresponding to tachyon mass $\mu_t = 4.3$ TeV.
Because $E_W \gg M_W$,
use of the equivalence theorem is probably
justified for these parameters.

To obtain the event rate for the $O(4)\times O(8)$ model
with parameters $m_\psi=125$ GeV and $\mu_t = 4.3$ TeV,
we fold the amplitudes with the parton luminosities.
We find that the elastic $W^-W^+$ event rate
for $M_{WW} \ge 350$ GeV at the SSC
(with $\sqrt{S}=40$ TeV
and integrated luminosity $10^4$ ${\rm pb}^{-1}$)
is about 0.5 pb.
This rate is about the same as the total rate (0.6 pb)
for the $O(4)$ model (\ie, the $n=0$ limit)
with a resonance with $m= 845$ GeV and $\Gamma =  640 $ GeV,
as expected from the scaling argument given above.
Moreover, this rate is not much smaller
than the rate (3.4 pb) for a 500 GeV standard model Higgs boson,
with width 64 GeV, produced via the $W$--fusion process.
The $Z^0 Z^0$ event rate in the $O(4) \times O(8)$ model
is about half the $W^+ W^-$ event rate.
We have not included in these rates $W$ pairs
produced by either quark or gluon fusion,
restricting our consideration to $WW$ scattering.

Many studies have been performed on detecting
Higgs bosons at the SSC.
It has been shown that a $\sim 500$ GeV standard model Higgs boson
can be detected using the ``gold--plated'' mode alone,
and does not require
the application of techniques such as jet--tagging\Ref\tag{
R.~N.~Cahn {\it et al.}, Phys.\ Rev.\ {\bf D35}, 1626 (1987);\nextline
V.~Barger, T.~Han, and R.~J.~N.~Phillips, Phys.\ Rev.\ {\bf D37}, 2005 (1988);
\nextline
R.~Kleiss and W.~J.~Stirling, Phys.\ Lett.\ {\bf 200B}, 193 (1988);\nextline
U.~Baur and E.~W.~N.~Glover, Nucl.\ Phys.\ {\bf B347}, 12 (1990);\nextline
D.~Dicus, J.~Gunion, L.~Orr, and R.~Vega, preprint UCD--91--10.}
and/or jet--vetoing\Ref\veto{
V.~Barger, K.~Cheung, T.~Han, and R.~J.~N.~Phillips,
Phys.~Rev. {\bf D42}, 3052 (1990);
D.~Dicus, J.~Gunion, and R.~Vega, Phys.~Lett.~{\bf B258},  475 (1991). }
used in studying TeV $WW$ interactions.
However, these techniques, together with others,
such as measuring the charged particle
multiplicity of the event\Ref\multi{
J.~F.~Gunion, G.~L.~Kane, H.~F.-W.~Sadrozinski, A.~Seiden, A.~J.~Weinstein,
and C.--P.~Yuan, Phys. Rev. {\bf D40}, 2223 (1989).}
or testing the fraction of longitudinal $W$'s,\refmark{\kane}
could be used to further improve the signal--to--background ratio
to study a $\sim 500$ GeV standard model Higgs boson
produced via $WW$ fusion processes.
We think it is clear that similar strategies could be applied
to detect the $\sim 500$ GeV resonance
in the $O(4) \times O(8)$ model
discussed above.

For the $O(4) \times O(32)$ model of ref.~2,
with parameters $m_\psi=125$ GeV and $\mu_t=2.5$ TeV,
the resonance ($m=275$ GeV and $\Gamma=120$ GeV)
is in a region where the energy $E_W$ of the longitudinal $W$
in the $WW$ center--of--mass frame
is less than twice $M_W$.\Ref\neqv{
The equivalence theorem,
which allows us to identify the Goldstone bosons $\phi$
with the longitudinal gauge bosons $W$
in the $WW$ scattering processes,
requires $E_W \gg M_W$.}
To see whether such a resonance could be detected
would require a detailed Monte Carlo study,
which we will not perform in this paper.
We would argue, however,
that with
$\sim 5000$ extra longitudinal $W^+W^-$ pairs and
$\sim 2500$ extra longitudinal $Z^0Z^0$ pairs
produced in one standard SSC year,
this signal could probably be observed
with appropriate detectors.

In this note,
we have considered the $O(4)\times O(n)$
model presented in ref.~\CG~containing
many inelastic channels
in the $WW$ scattering process.
We demonstrated through a simple scaling argument that,
although the amplitude for elastic $WW$ scattering
in this model decreases
as the number $n$ of inelastic channels increases,
the total elastic event rate remains more or less the same.
(We choose the parameters for each model
to give the ``heaviest'' possible resonance.)
This rate is about the same as that
for the $O(4)$ model,
with no inelastic channels and a heavy resonance.

\ack

We would like to thank G.~L.~Kane
for asking questions which stimulated this work,
and for insisting that the total elastic cross--section
could not decrease with many inelastic channels.
We are also grateful to
J.~Bagger, Gordon Feldman, C.~Im,
G.~Ladinsky, S.~Meshkov, F.~Paige,
E.~Poppitz, and E.~Wang for discussions.
This work has been supported by the National Science Foundation
under grant no.~PHY-90-96198.

\refout

\end